# High speed 3-D Surface Profilometry Employing Trapezoidal HSI Phase Shifting Method with Multi-band Calibration for Colour Surface Reconstruction

**L C Chen, X L Nguyen and Y S Shu**
National Taipei University of Technology, Taipei, Taiwan

Email: lcchen@ntut.edu.tw

**Abstract.** This article presents a new optical measurement method employing a HSI (Hue, Saturation and Intensity) colour model to form trapezoidal structured patterns for morphology reconstruction of a measured object at a high speed. Profilometry on objects having non-monochromatic surfaces is considered as one of the remaining most challenges faced by the currently existing structured-light projection methods since the surface reflectivity to red, green and blue light may vary significantly. To address this, an innovative colour calibration method for hue component is developed to determine the accurate reflectivity response of the measured surface. The trapezoidal colour pattern is calibrated to compensate the hue-shifted quantity induced by the reflective characteristics of the object's surface. The developed method can reconstruct precise 3-D surface models from objects by acquiring a single-shot image, which can achieve high-speed profilometry and avoid *in-situ* potential measurement disturbances such as environmental vibration. To verify the feasibility of the developed methodology, some experiments were conducted to confirm that the measurement accuracy can be controlled within 2.5% of the overall measurement range and the repeatability of 3.0% within ±3σ can be achieved.

## 1. Introduction

The importance of high-speed 3-D surface profilometry has been highlighted by many in-field applications of modern metrology technology worldwide. A considerable amount of research and development work has been performed to increase the performance of 3-D measurement systems using automatic optical inspection (AOI) techniques. With most industrial *in-situ* measurement systems, vibration disturbance is one of the main significant issues to be overcome in obtaining accurate 3-D information, especially in *in-situ* process inspection environment. Strategies to avoid the use of expensive anti-vibration facilities or strict operation requirements for in-field 3-D measurement have thus become a critical issue. Speeding up the acquisition process is critical for achieving real-time measurement. The traditional phase-shifting techniques [1] employed in optical interferometry, moiré, and fringe projection methods for 3-D surface profilometry have been widely studied for decades. These techniques mainly focused on measurement accuracy. However, due to the use of multiple fringe images required for phase calculation, the existing methods are prone to troublesome measurement errors caused by unavoidable environmental vibration or other external disturbances.

Applied to real-time issues in industrial applications, Fourier transform profilometry (FTP) has emerged as an important method of 3-D profilometry, initially proposed by Takeda [5]. One shot imaging capable of 3-D surface reconstruction is the main advantage of this method. In its development, the FTP method has been improved from the original 1-D Fourier Transform with single-frequency fringes to 2-D Fourier Transform with multiple-frequency fringes [6]. Other improvements include using Fourier Transform speckle to increase measurable step heights [7], modifying spectral filters to obtain better measurement performance [8], and encoding patterns of tilting-fringe projection [9]. The challenge for FTP is to effectively separate first-order spectral data from the other frequency components, so that 3-D surface reconstruction can be accurately achieved. This issue also reduces the accuracy of the FTP method, so its results are generally not as good as those obtained by the phase-shifting technique.

In contrast to the FTP method, techniques employing colour encoded patterns to obtain simultaneous multiple phase shifting information in one projected pattern can be divided into two major research directions, multiple-colour stripe projection and colour phase-shifting technique. The multiple-colour stripe projection employs a structured pattern encoded by multiple colour stripes to be projected onto the object's surface [10]. Each stripe has its unique colour features for further retrieving the phase information with respect to the surface height in its decoding process. Adaptive strategies of colour projection have been applied to reduce the effect of scene characteristics [13]. This kind of technique is generally simple to implement and relatively fast in its acquisition process. However, to achieve a high spatial resolution in measurement, the number of stripes must be significantly increased, which is not always feasible due to the limited number of projecting stripe colours available for effective coding. In addition, increasing the number of the encoded colours can also raise undesired noise from the uncertain surface colour reflection and the complexity of the coding and decoding processes. In contrast to the above method, the colour phase-shifting technique is primarily based on the principle of the traditional phase-shifting method and simultaneous multiple colour fringe encoding. According to the important assumption of adequate surface colour light reflection to be acquired for each projected colour fringe, the three colour (namely red, green and blue) channels of the deformed fringe image can be effectively acquired and split into three single colour phase-shifted patterns to achieve an instantaneous three-step phase shifting operation [15]. Those methods have the problem of separating the three color channels since color cross talk phenomenon commonly exists in sensor detection. To resolve this, a blind color isolation (BCI) algorithm was developed to resolve the above issue by determining a demixing matrix [20]. The main advantage of this technique is that it can obtain phase-shifting patterns simultaneously and therefore can achieve one-shot 3-D surface profilometry. However, as cited in the previous literature, in order to accurately establish three independent phase shifting information, the method must overcome a non-trivial difficulty encountered in potential light coupling and ambiguity problems of the three colour channels. Until now, the problem has not been effectively addressed, so that the accuracy of the 3-D surface reconstruction is not strictly guaranteed.

To overcome this problem, this article presents a new surface profilometry approach using a new concept of employing the HSI colour model with trapezoidal fringe pattern projection to achieve one-shot 3-D surface profilometry with high measurement accuracy. The trapezoidal HSI colour model has been studied recently [21]. However, the method can be only performed on white surfaces since colour reflectivity uncertainty still remains one of the most difficult issues to battle. To obtain the reflectivity characteristics of the object's surface, a colour calibration method on the hue component is developed here to compensate possible hue distortion induced by unexpected surface reflectivity detected from various measured surfaces. The hue component determined by the developed method is immune to potential variance of surface colour reflectivity, which potentially induces unacceptable measured errors using the existing phase shifting methods. With the proposed system calibration method for correction of non-linear colour distortion, the method can reconstruct 3-D surface maps accurately with a measurement speed up to the frame rate of the image-sensing unit.

The rest of the paper is organized as follows. Section 2 describes the methodology and calibration of 3-D surface profilometry using the HSI colour transformation and trapezoidal phase-shifting algorithm. To perform 3-D surface measurement, Section 3 presents the optical system design and analyzes the performance of the developed methods for 3-D surface reconstruction. Section 4 summarizes the study.

## 2. Methodology

The schematic diagram and the flow chart of the proposed method are shown in Figure 1 and Figure 2, respectively. The proposed method generates the trapezoidal colour fringe patterns by combining three-colour phase-shifted fringe patterns into a single structured pattern in which the period of projecting fringes can be controlled flexibly. After the pattern is projected on to the measured surface, the colour image captured by a triple-CCD camera, in which each image are acquired by three colour

CCDs (Red, Green and Blue channels) independently, is transformed into the HSI model where the hue component is extracted as the phase quantity in phase-shifting method. Then, the hue component is further compensated by a mapping process using a lookup table. The phase unwrapping process is then applied to obtain the continuous phase map of the image. By using triangulation, the relation between the hue value and the height information of object's surface can be determined.

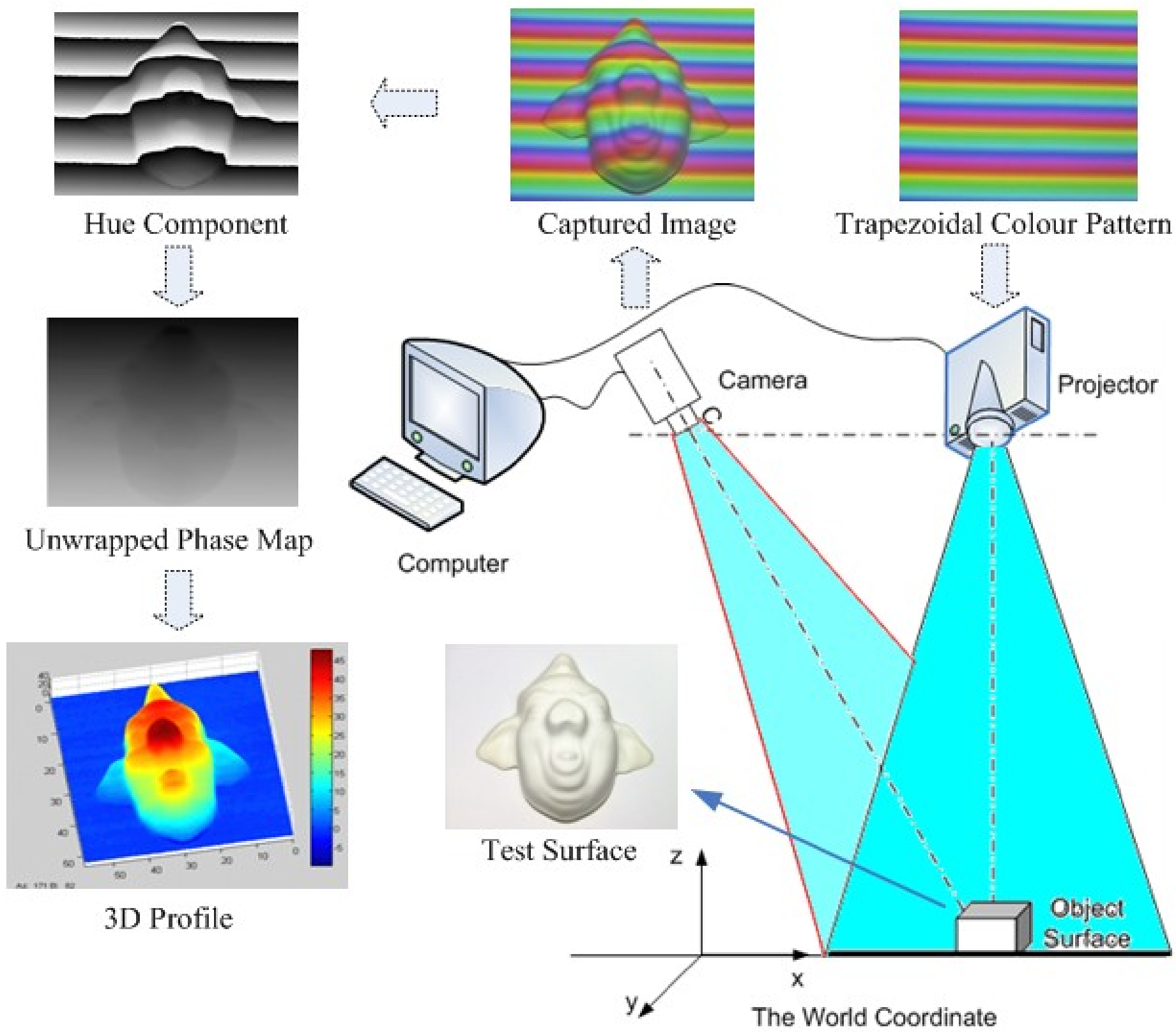

**Figure 1.** Schematic diagram of the proposed 3-D surface profilometer.

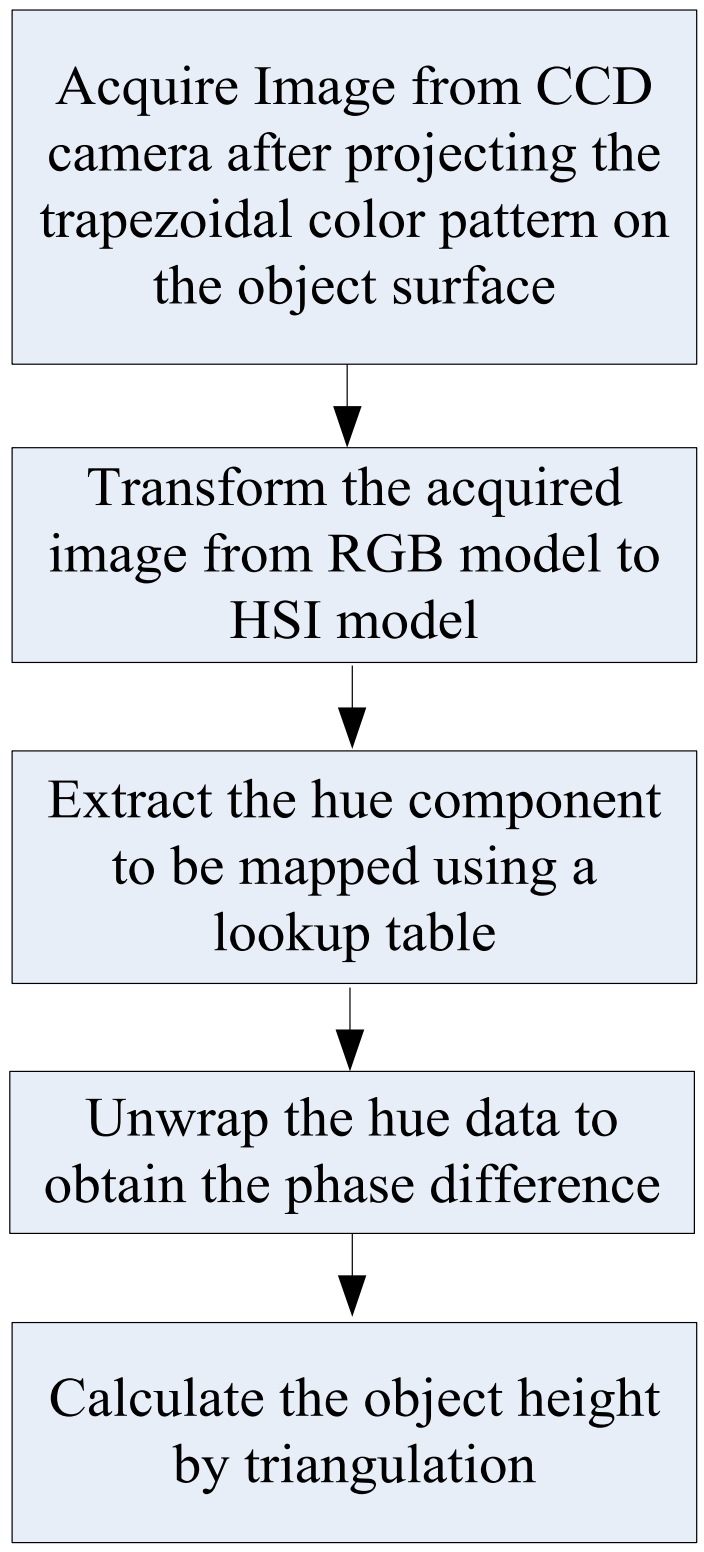

**Figure 2.** Flow chart of the proposed measurement method.

### 2.2. RGB to HSI colour model transformation

The deformed fringe image captured from the triple-CCD camera is transformed into an HSI model to extract the hue component. The hue component of a point having three colour components of ($I_r$, $I_g$, $I_b$) in an image can be expressed as follows [22]:

$$H = \cos^{-1}\left\{ \frac{\frac{1}{2}\left[(I_r - I_g) + (I_r - I_b)\right]}{\left[(I_r - I_g)^2 + (I_r - I_b)(I_g - I_b)\right]^{1/2}} \right\} \tag{1}$$

where $I_r$, $I_g$, and $I_b$ are the light intensities of red, green and blue, which have been normalized to a range of [0, 1].

Equation 1 yields the value of H in an interval of $0 \le H \le \pi$. To obtain the hue value in $\left[0, 2\pi\right]$, the following transformation is considered:

$$H = \begin{cases} H, & \text{if} \quad I_b < I_g \ ; \\ 2\pi - H, & \text{if} \quad I_b \ge I_g . \end{cases} \tag{2}$$

The trapezoidal colour fringe pattern with a constant value of RGB component in the vertical direction y is designed as follows:

$$I_r(x,y)=\begin{cases}1 & x\in[0,T/6)\ or\ [5T/6,T)\\(2-6x/T) & x\in[T/6,T/3)\\0 & x\in[T/3,2T/3)\\(6x/T-4); & x\in[2T/3,5T/6)\end{cases}$$

$$I_g(x,y)=\begin{cases}(6x/T) & x\in[0,T/6)\\1 & x\in[T/6,T/2)\\(4-6x/T) & x\in[T/2,2T/3)\\0 & x\in[2T/3,T)\end{cases}$$

$$I_b(x,y)=\begin{cases}0 & x\in[0,T/3)\\(6x/T-2) & x\in[T/3,T/2)\\1 & x\in[T/2,5T/6)\\(6-6x/T) & x\in[5T/6,T)\end{cases} \tag{3}$$

To obtain a continuous and linear phase in the projected fringe pattern, it is important that the hue component is linear after being transformed from its trapezoidal RGB pattern to the HSI model, as shown in Figure 3. Six sections are included in each period of the encoded structured light pattern, in which each individual spatial length equals to T/6 and T represents for one spatial period of the projected colour fringe pattern.

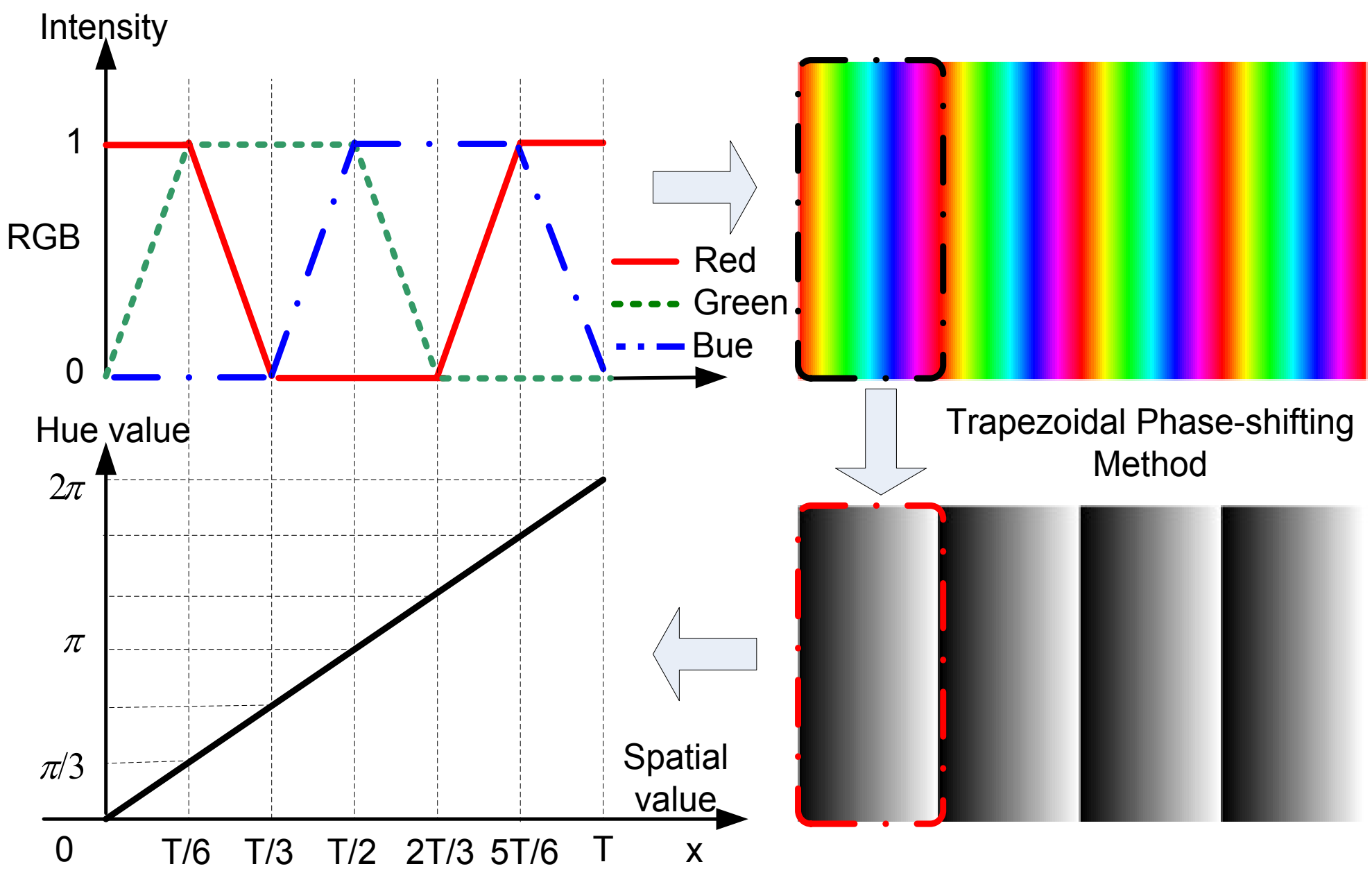

**Figure 3.** Hue component to be calculated from the trapezoidal colour fringe pattern (RGB model).

Since these sections are symmetric and similar, the first section is first analyzed and its result can be applied to the other sections. In the first section, $x\in[0,T/6)$, in which $I_r=1$; $I_g=6x/T$; $I_b=0$, the hue component can be expressed as:

$$H=\cos^{-1}\left\{\frac{\frac{1}{2}\left[\left(1-\frac{6x}{T}\right)+1\right]}{\left[\left(1-\frac{6x}{T}\right)^2+\frac{6x}{T}\right]^{\frac{1}{2}}}\right\}=\cos^{-1}\left\{\frac{1-\frac{\left(\frac{6x}{T}\right)}{2}}{\left[\left(\frac{6x}{T}\right)^2-\frac{6x}{T}+1\right]^{\frac{1}{2}}}\right\}\approx\cos^{-1}\left\{1-\frac{\left(\frac{\pi}{3}\frac{6x}{T}\right)^2}{2}\right\} \tag{4}$$

for $x\in[0,T/6)$.

Since 1 and $\frac{1}{2}\left(\frac{\pi}{3}\frac{6x}{T}\right)^2$ are the first and second components in the Taylor series of $\cos(\frac{\pi}{3}\frac{6x}{T})$, Equation (4) can be simplified as:

$$H\approx\frac{\pi}{3}\frac{6x}{T}=Cx \tag{5}$$

This is an approximately linear relationship between the hue component and the spatial coordinate of x-axis, where $C$ is a constant coefficient. The errors of this non-linear approximation can be further eliminated by using a lookup table for mapping an accurate hue value. The linear relationship between the hue component and the spatial coordinate of x-axis can be also generated by keeping saturation and intensity components constant and using hue with the type of ramp signal to produce a RGB pattern [23]. The method could work in a manner similar to the proposed method; although it has a drawback that the RGB components in the generated pattern may contain some undesired sudden colour changes, which decrease the fringe contrast and measurement accuracy. The proposed method employs a trapezoidal colour pattern in which the red, green and blue channels of the projected pattern are all linear and symmetric along the spatial space. This can ensure a good fringe contrast of the colour pattern projection to be obtained.

### 2.3 System calibration method

Measurement using a structured light projection method could be affected by many factors including the gamma effect in the light projector, surface reflectivity of the test object and nonlinear sensitivities for three colour channels of the triple-CCD camera. When using the colour-encoded pattern projection method, surface profilometry on a test object having various surface colours usually encounters unacceptable measurement errors since the light reflection of the surface may vary significantly against each colour component of the projected light. The linear hue of the encoded structured pattern can be distorted by three various kinds of non-linearity when it is transferred through the light propagation process illustrated in Figure 4. This undesired condition can significantly reduce measurement accuracy of the developed method. The intensity of light received by the image sensor is transformed through three stages in its light propagation process, as explained below.

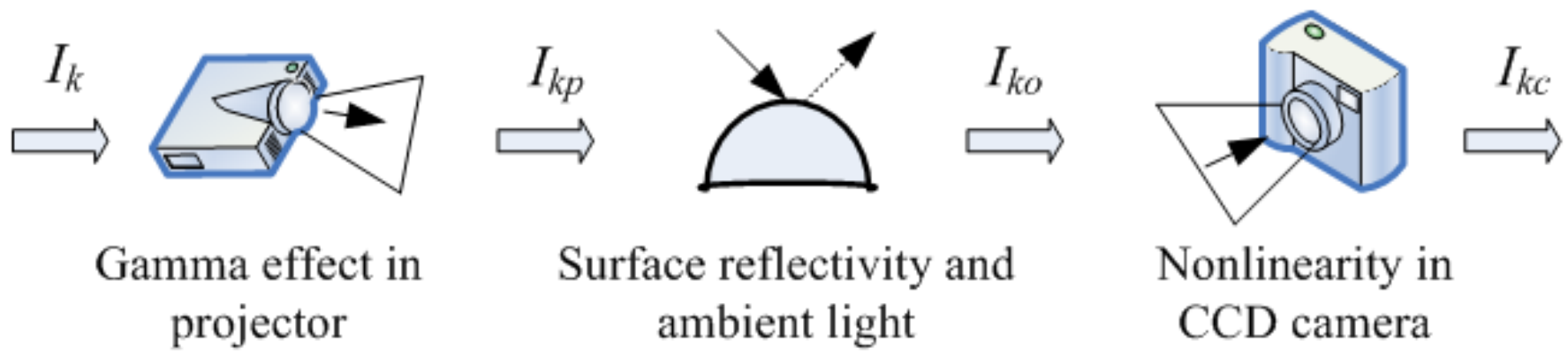

**Figure 4.** Potential distortion of the hue model due to various non-linear conditions encountered in the structured light transformation process.

Assuming that the trapezoidal structured light pattern is encoded by the three color intensities, $I_k(x,y)$ with *k*= *r,g,b* representing three fundamental colour channels, the intensities of the fringe images generated by the light projector can be described as follows:

$$I_{kp}(x,y) = g_k(I_k(x,y)) \tag{6}$$

Where $g_k(I_k)$ is the gamma function of the projector for the colour channel *k*.

Furthermore, the intensity of the reflected light from the test object's surface can be affected by surface reflectivity, $r_k(x,y)$, and the ambient light, $a_k(x,y)$, from the surrounding environment. Due to various reflectivities of the object surface, each channel of the projected structured light can possess different coefficients of the light reflective response. In the research, it is assumed that the three color channels can receive its individual reflected light with an adequate level of signal to noise (S/N) ratio. The structured light pattern upon contacting with the object surface can be described as:

$$I_{ko}(x,y) = r_k(x,y)\left[I_{kp}(x,y) + a_k(x,y)\right] \tag{7}$$

The reflected light receiving from the object surface is further acquired by the triple-CCD camera. In general, the three chip may possess different sensor sensitivities, $\alpha_k$ (*k*= *r,g,b*), for the three colour channels. Thus, the light intensities of fringe pattern captured by the CCD camera can be expressed:

$$I_{kc}(x,y) = \alpha_k\left[I_{ko}(x,y) + b_k(x,y)\right] \tag{8}$$

where $b_k(x,y)$ is the ambient light that comes from the environment directly and is further received by the triple-CCD camera.

Therefore, the captured hue component may not be as ideal as given in Equation (1) and it can be described as a function of all the above parameters:

$$H = f\left(I_k, g_k, r_k, \alpha_k, b_k\right) \quad \text{with } k = r,g,b \tag{9}$$

To address this potential problem effectively, the proposed method employs a developed colour calibration method (described in Figure 5 and 6) to find the above system parameters. Assuming that the measurement is performed within a controlled measurement environment, the ambient light is minimized to an insignificant level. In addition, it is also assumed that triple-CCD camera is under a cooling temperature control, so its dark current level is controlled within a stable and low level. The system parameters of the projector and the triple-CCD camera are kept as a fixed set of measurement parameters after they are set at an adequate level.

In the calibration process, a white-reference structured pattern with a light intensity (255,255,255) of three colour channels is projected onto the object's surface and then captured by the triple-CCD camera for performing a white balance operation on the projector. Using the above procedure, the parameters used in the projector are adjusted according to the normalized values of the red, green and blue colour channels. Then 256 uniform patterns of different hue values ranging equally from 0 to $2\pi$ are generated and projected to the measured surface. These sequential colour images are captured by the triple-CCD camera and then transformed to the HSI domain. This procedure is repeated three times to reduce the effect of random noise. For each projected hue value, an interpolation process is applied to find the corresponding hue from three captured hue values which may be influenced by the actual reflectivity response of a measured surface.

By performing the above process, a lookup table describing the colour reflectivity characteristics of the measured surface can be established at every pixel of the captured image. This table provides crucial information to compensate for the projected trapezoidal colour pattern, so that the hue component can maintain a linear relationship with the detected height. The compensated value for each colour channel (red, green or blue) is initialized with the unit value and can be positive or negative according to whether the captured hue is higher or lower in the colour wheel. After the compensation process is completed, the two sets of projected and captured hues are further compared to determine the standard deviation. If the standard deviation is less than a preset threshold, the calibration process is terminated and measurement of the object's surface can be performed. Otherwise, the compensated value is increased and the calibration process resumes for another try until the deviation is converged satisfactorily. With the developed calibration process, the calibrated compensating value of three projected colours for each projected surface pixel can be employed to eliminate potential deviation caused by various surface colour reflectances.

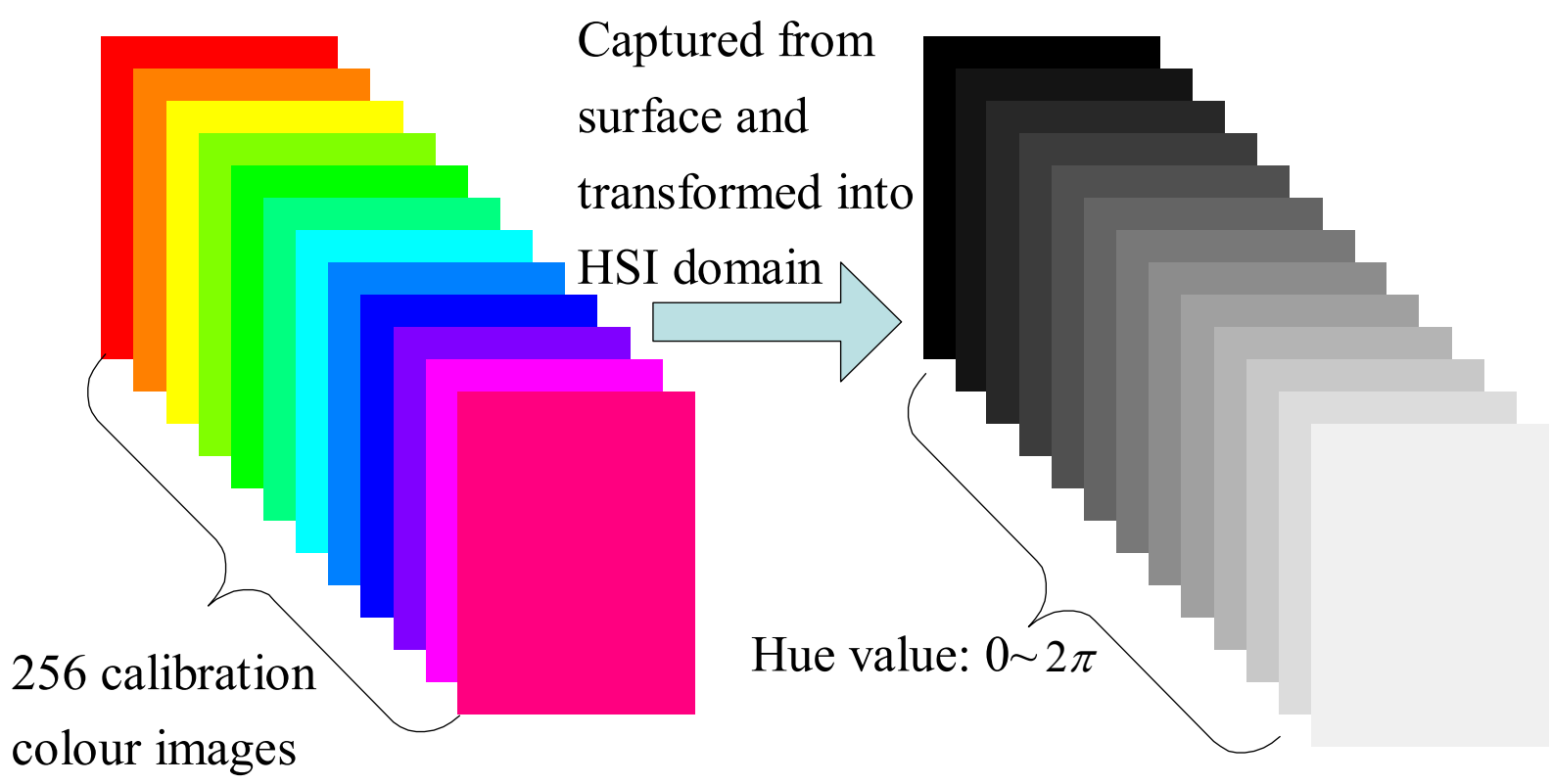

**Figure 5.** Projection of multiple uniform colour images to determine the surface reflectivity response.

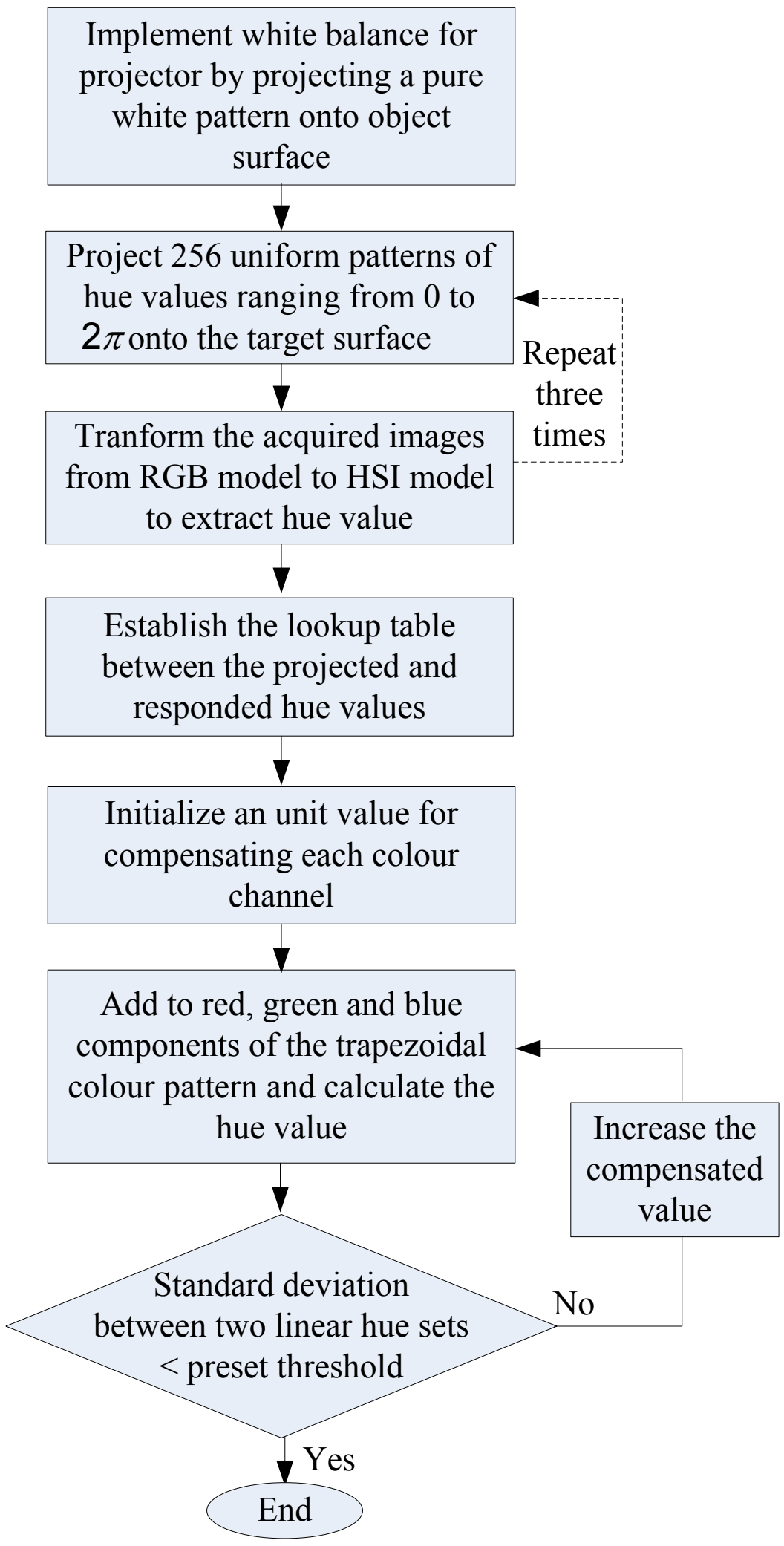

**Figure 6.** Flow chart of the colour calibration process.

The measurement process without and with the colour calibration are compared in Figure 7, where a calibrated flat board with a surface flatness less than 0.2 μm is measured. For the case without the colour calibration, the phase map of the flat board appears with ripples (shown in Figure 7(c)) that are accumulated by many effects either from the projector, object's surface or camera. These effects are effectively resolved by applying the developed colour calibration process which feeds back to change the produced trapezoidal colour pattern. Shown in Figure 7(d), the desired phase map with linear hue component is accurately obtained.

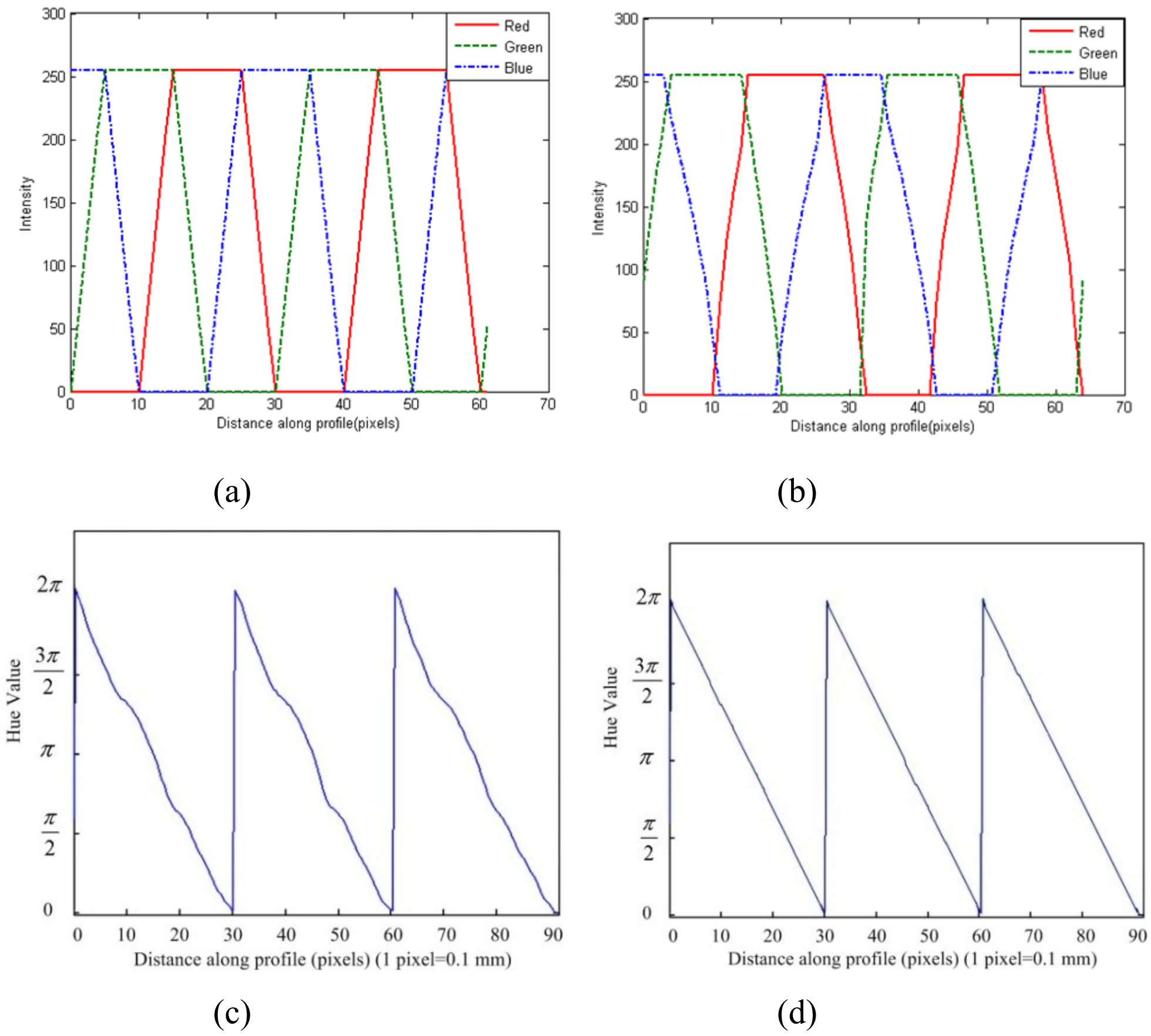

**Figure 7.** Measurement analysis of a calibrated flat board: (a) and (b) show the encoded patterns to be projected on a calibrated flat board using a Digital Light Projector (DLP) without and with colour calibration process, respectively; (c) and (d) show the cross sections of the captured hue without and with the colour calibration process, respectively.

## 3. Experiment results

The hardware setup of the developed system is shown in Figure 8. It consists of a digital micromirror device (DMD) having a 1024 × 768 pixel resolution for generating colour encoded fringe patterns, a triple-CCD (Sony DXC-390) camera with a high speed of up to 60 fps, a personal computer equipped with a dual core Intel Pentium D having a speed of 3400 MHz with 1 GB SDRAM for controlling the projector and acquiring the images through a suitable frame grabber and a set of optical lenses for obtaining the desired characteristics of the fringe pattern.

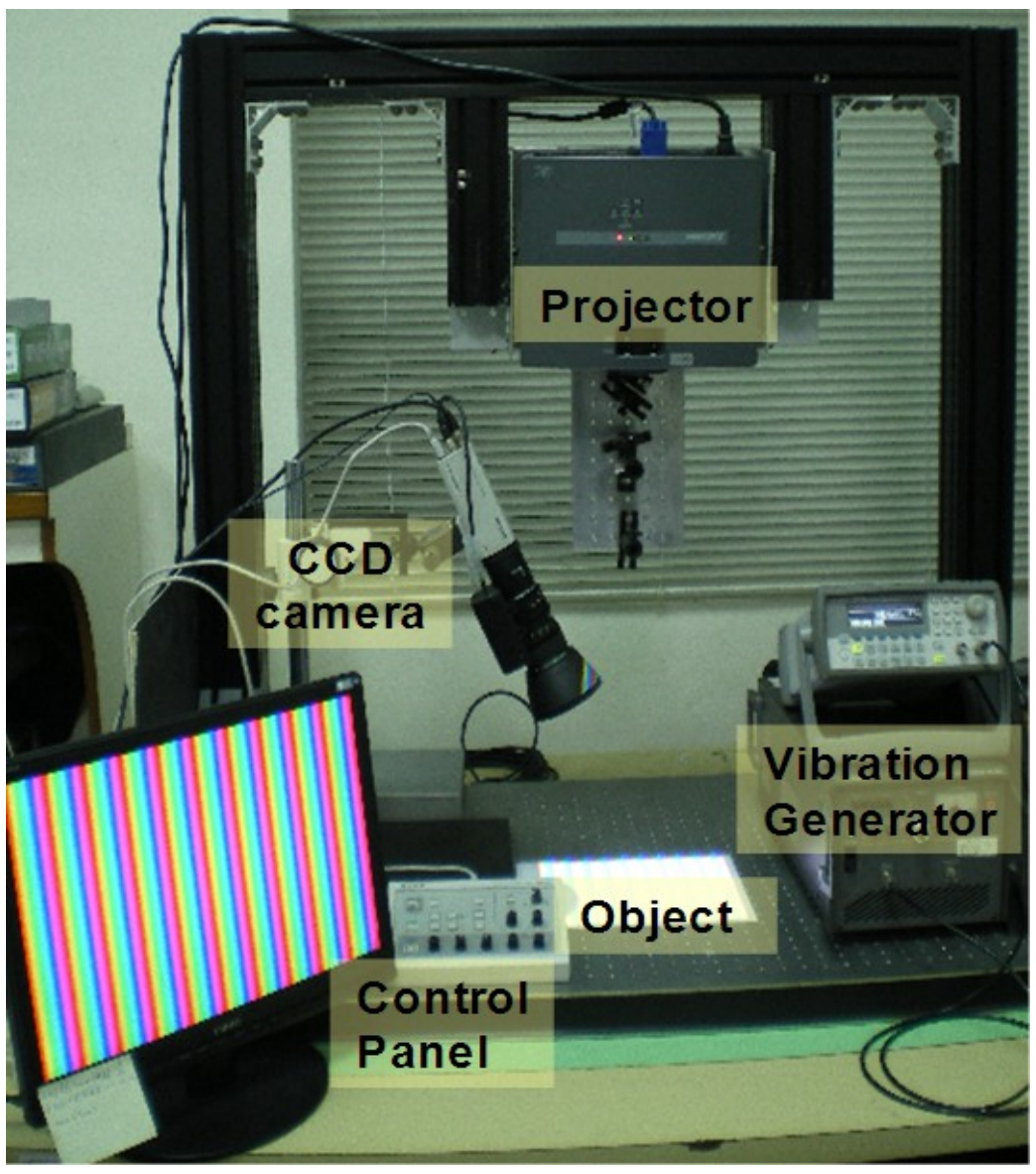

**Figure 8.** System hardware setup for the developed methodology.

*3.1. Industrial sample measurement*

In this experiment, a set of colour-coated ceramic gauge blocks having three calibrated step heights of 3.0 ± 0.001 mm (green), 1.0 ± 0.001 mm (blue) and 4.0 ± 0.001 mm (red), respectively, being mounted on a vibratory membrane with controllable frequencies, were measured to verify the accuracy of the developed method in measuring vibratory colour objects and validate its one-shot imaging performance. The measured surface of the gauge blocks was of ceramic type being coated with uniform red, green and blue thin-films on each step height, in which its surface flatness was verified to be superior to 30 nm. The experiment with gauge blocks was conducted under a fluctuating frequency of the excited membrane ranging from 0 to 60 Hz. By using the proposed calibration method, the projecting fringe pattern was adjusted according to the surface colour coated on the three gauge blocks, in order to maintain the linearity between hue and spatial value. Results of the experiment are shown in Figure 9. A general least-square evaluation method confirmed that measurement accuracy of the three step heights were controlled within a maximum deviation of 2.5% of the overall measuring depth range. The acquisition time of approximate 3.5 ms and a 3-D map detection rate of 60 frames per second (fps) were achieved for one-shot 3-D surface profilometry.

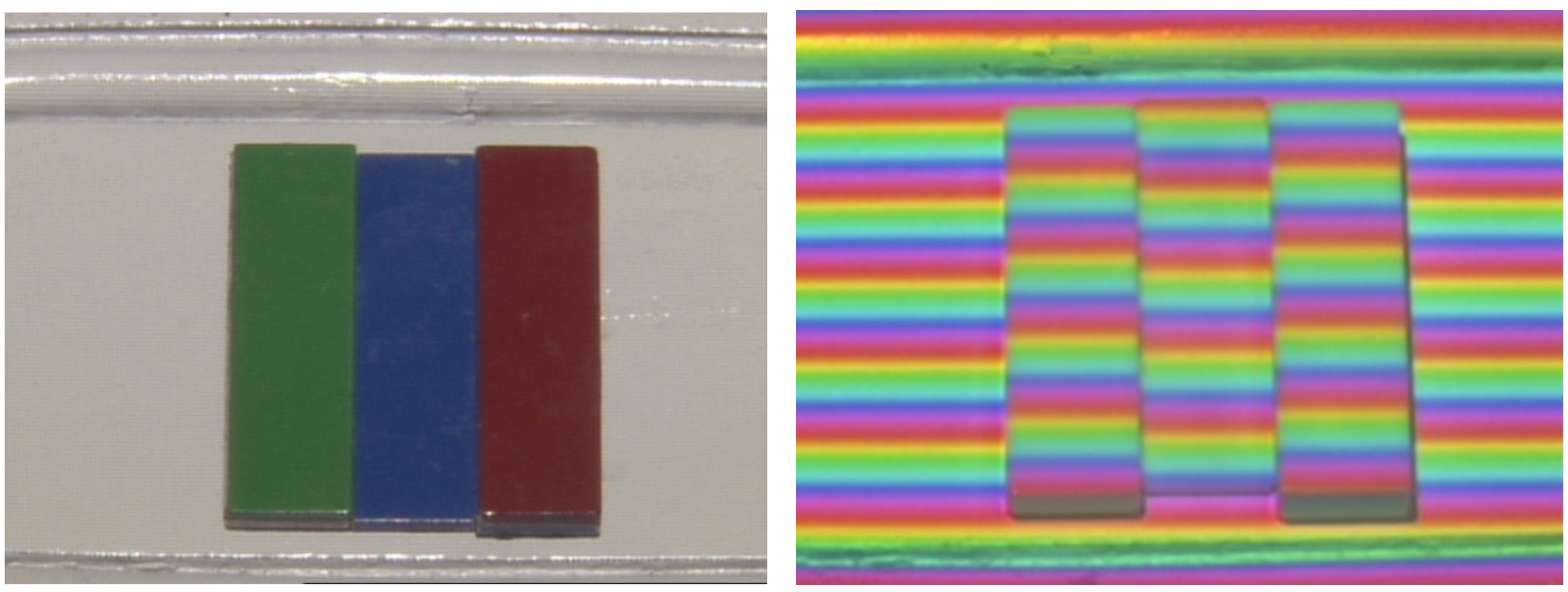

(a) (b)

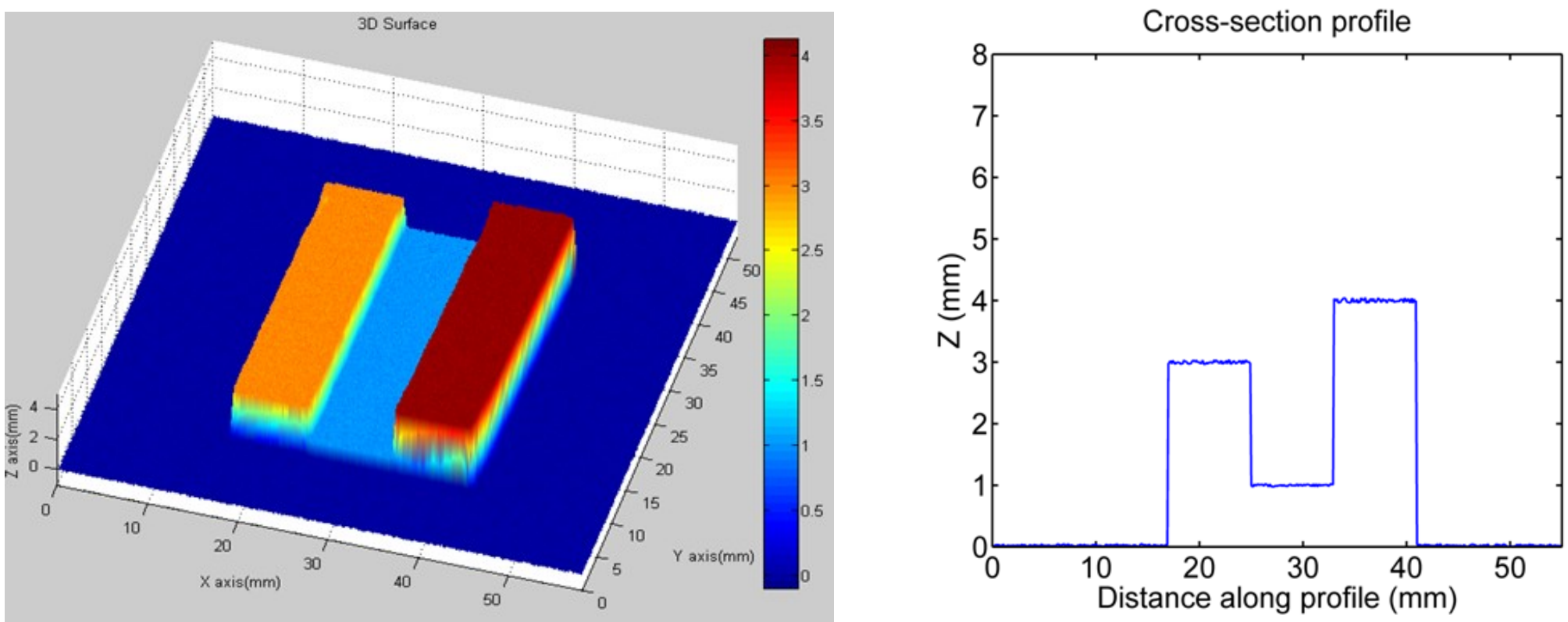

(c) (d)

**Figure 9.** Measurement results of the tested step-heights at frequency of 30Hz: (a) Gauge blocks mounted on a vibratory membrane; (b) the deformed fringe image; (c) the 3-D profile; (d) Cross-section profile of three gauge blocks surface.

Another experiment is the measurement of a human dental model to demonstrate the capability of proposed method for measuring objects with complex structure and non-monochromatic surface (shown in Figure 10(a)). By colour calibration, the structured colour pattern can be adjusted to produce a true hue map for the dental surface. The trapezoidal colour pattern was projected onto a measured surface, from which the phase map can be obtained (Figure 10(b) and Figure 10(c)). The 3-D profile is shown in Figure 10(d), (e) and (f). It is confirmed that the developed system is capable of providing accurate 3-D surface measurement for many practical applications.

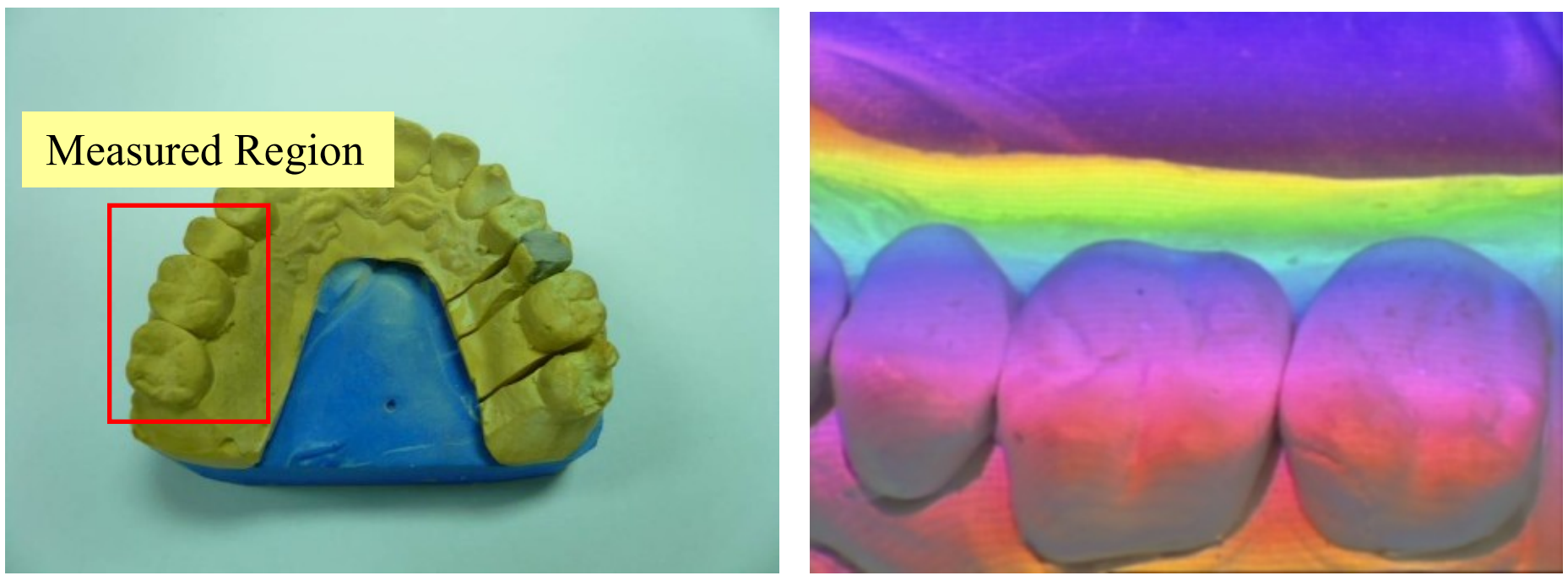

(a) (b)

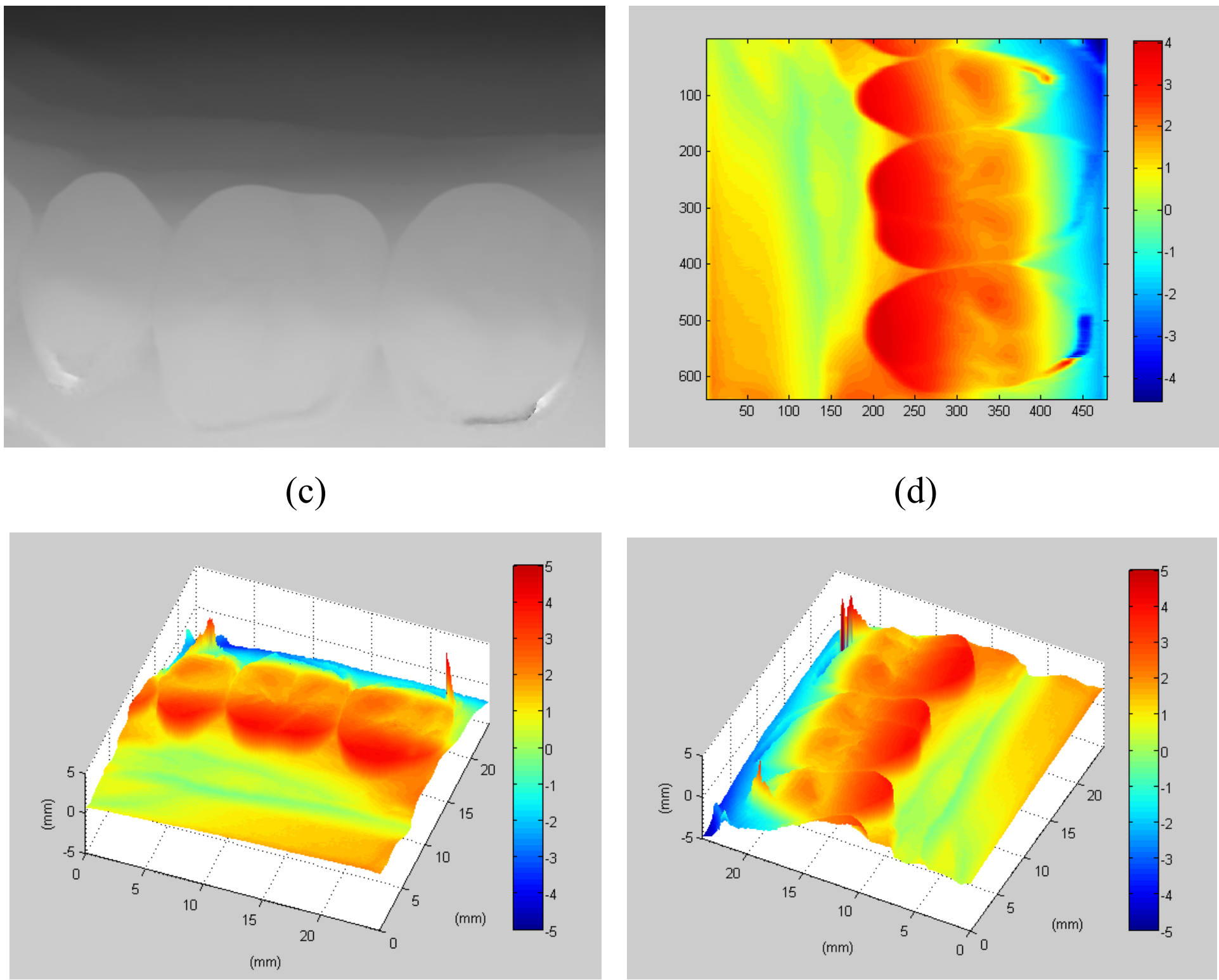

(c) (d)

(e) (f)

**Figure 10.** Human dental model measurement results: (a) human dental model; (b) deformed fringe image; (c) phase map; (d) two dimensional image with height distribution; (e) and (f) different views of the 3-D profile.

*3.2. Comparison with traditional three-step phase-shifting method*

The research focuses on *in-situ* applications of automatic optical profilometry, which is related to many modern industrial manufacturing processes. The traditional phase-shifting method is widely used since it can provide high accuracy. However, vibration disturbance resistance is one of the main weaknesses of this method. The proposed method using one-shot RGB technique can overcome the vibration disturbance issue encountered by multiple-step phase shifting principle. To demonstrate its efficiency and precision, the proposed method was compared with the general three-step phase-shifting method using the developed system. A calibrated blue-coated ceramic gauge block with standard height of 1.0 ± 0.001 mm in the first experiment (section 3.1) was employed to evaluate the measurement accuracy. The imaging pixel pitch was maintained at 0.2 mm for both sets of measurements. Each method was conducted by five experiments using different fringe periods of 20, 40, 60, 80 and 100 pixels, equivalent to 4, 8, 12, 16, 20 mm in the spatial domain of the measurement reference plane, respectively. For each experiment with a fringe period, the blue-coated ceramic gauge block was measured thirty times for calculating standard deviation and average error. Results of the proposed method and the traditional three-step phase-shifting method are shown in Table 1. From a general evaluation, the experiments show that the proposed method can provide higher accuracy than the traditional phase-shifting method with the same conditions of measurement and fringe period.

**Table 1.** Measurement results of the colour-coated gauge block (1.0± 0.001 mm) by the proposed method and by traditional three-step phase-shifting method

| Number of pixels in one fringe period (1pixels=0.2mm) | Proposed method | | Traditional three-step phase-shifting method | |
|---|---|---|---|---|
| | σ(standard deviation) (%) | Average error $\overline{e}$ (%) | σ(standard deviation) (%) | Average error $\overline{e}$ (%) |
| 100 | 1.29 | 1.05 | 3.69 | 3.17 |
| 80 | 1.33 | 1.2 | 3.72 | 3.14 |
| 60 | 1.38 | 1.28 | 3.78 | 3.25 |
| 40 | 1.57 | 1.69 | 3.85 | 3.34 |
| 20 | 1.94 | 1.98 | 3.93 | 3.37 |

## 4. Conclusions

The trapezoidal colour pattern projection using the developed HSI model only requires one-shot imaging to reconstruct 3-D profiles of an object having non-monochromatic surfaces. The 3-D measurement method can overcome potential fluctuating light intensity and unwanted noise encountered in optical measurement processes. The proposed trapezoidal colour fringe pattern with simultaneous three colour-encoded fringes has unique advantages for one-shot high speed 3-D surface profilometry, especially effective for *in-situ* AOI. With the development, a 3-D detection rate up to 60 fps or higher can be achieved and the measurement accuracy can be maintained within a maximum of 2.5% variation of the total detection range. The experimental results have verified the feasibility of this method for real-time dynamic 3-D measurement with bandwidths up to 60 Hz or higher. The proposed method is suitable for measuring objects having single colour surface. For objects having arbitrary colour surface, a further calibration technique needs to be developed, which can be addressed in future works.

## Acknowledgments

The authors would like to thank the National Science Council (NSC) of Taiwan, for financially supporting this research under Grant NSC 96-2628-E-027-005-MY3.

## Reply to the reviewer 1's report

The authors would like to thank the reviewer for the suggestions and valuable comments, which help improve much the paper. I sincerely appreciate the reviewer's great efforts in pointing out the inconsistencies and errors. The manuscript has been revised according to the reviewer's comments.

**[Comment and Reply]**

(1) In Fig. 7, what is the meaning of DLP? There is no explanation across the whole paper. Why the authors compare the encoded patterns by DLP with the captured hue? I think the hue value is corresponding to the phase value. So the comparison should be between the hue value and phase value instead of intensity.

[Reply]

We are sorry for missing the explanation of DLP, which stands for "Digital Light Projector". We have added the explanation in Fig. 7.

The encoded patterns by DLP in Fig. 7(a) and Fig. 7(b) are given just for reference. Yes, the hue value represents for the phase value and in this figure, as well as stated in the text, we would like to provide a comparison for hue values without and with colour calibration process in Fig. 7(c) and Fig. 7(d), respectively. So from these results, we can clearly see the effectiveness from applying the developed colour calibration process. The produced trapezoidal colour pattern resulted in Fig. 7(b) can be used to generate a linear hue distribution in Fig. 7(c), instead of the one in Fig. 7(d).

*In P.11, Line 12*

Change:

**Figure 7.** Measurement analysis of a calibrated flat board: (a) and (b) show the encoded patterns to be projected on a calibrated flat board using a DLP without and with colour calibration process, respectively; (c) and (d) show the cross sections of the captured hue without and with the colour calibration process, respectively.

To:

**Figure 7.** Measurement analysis of a calibrated flat board: (a) and (b) show the encoded patterns to be projected on a calibrated flat board using a Digital Light Projector (DLP) without and with colour calibration process, respectively; (c) and (d) show the cross sections of the captured hue without and with the colour calibration process, respectively.

(2) The authors argue that the experiment results verify that the proposed method can provide higher accuracy than three step phase shifting method with the same condition of measurement and fringe period. Then what is the condition of measurement and fringe periods in the experiments?

[Reply]

The proposed method and three step phase shifting method are employed to measure a standard height of 1.0 ± 0.001 mm by using the same developed hardware system setup. The environmental conditions such as room temperature and ambient light condition are kept the same. As stated in manuscript, the imaging pixel pitch was maintained at 0.2 mm for both sets of measurements. Each method was conducted by five experiments using different fringe periods of 20, 40, 60, 80 and 100 pixels, equivalent to 4, 8, 12, 16, 20 mm in the spatial domain of the measurement reference plane, respectively. For each experiment with a fringe period, the blue-coated ceramic gauge block was measured thirty times for calculating standard deviation and average error.

(3) To make the article more persuasive, it will be good to describe the background of the application. It is desirable to explain why the comparison is made between the proposed one and three step phase shifting method instead of one-shot RGB, since the proposed one is also one-shot.

[Reply]

Yes, our research focuses on *in-situ* applications of automatic optical profilometry which are related to many modern industrial manufacturing processes. The most popular method used in those systems is the phase-shifting method which can provide high accuracy, but *in-situ* vibration disturbance is one of the main issues influencing its effectiveness. So far in automatic optical inspection for industrial applications, the one-shot RGB method hasn't been widely applied due to its measurement repeatability. Therefore, we prefer to perform the comparison between the developed method and the aforementioned phase shifting method. By some experiments successfully conducted on measuring targets with colour surfaces, it is concluded that the proposed method can provide higher accuracy than the traditional phase-shifting method with the same conditions of measurement and fringe period.

To clarify this in the revised manuscript, the background of application has been added in the beginning of section 3.2 as follows:

*In P.14, Line 10*

Add:

"The research focuses on *in-situ* applications of automatic optical profilometry, which is related to many modern industrial manufacturing processes. The traditional phase-shifting method is widely used since it can give high accuracy. However, vibration disturbance resistance is one of the main weaknesses of this method. The proposed method using one-shot RGB technique can overcome the vibration disturbance issue encountered by multiple-step phase shifting principle."

## Reply to the reviewer 2's report

The authors would like to thank the reviewer for the suggestions and valuable comments, which help improve much the paper.

**[Comment and Reply]**

(1) I have just a small doubt on this phrase, that is not part of the article text, but it is contained in the author's reply (citing):
"The total time of the image acquiring and the phase calculation process was 156ms. This indicates that a 3-D detection rate can be achieved at up to 60 frame per second."
Is 6fps to be intended here? To achieve 60fps you need acquisition times <= 16.7ms
Just my curiosity.

[Reply]

Sorry for the mistyping we made in last reply. The sentence should be:
"The total time of the image acquiring and the phase calculation process was 15.6ms"
It means our system has the 3-D detection rate of 60 fps.